\documentclass[%
 aip,
 amsmath,amssymb,
 reprint,%
]{revtex4-1}

\usepackage{graphicx}
\usepackage{dcolumn}
\usepackage{bm}

\usepackage[utf8]{inputenc}
\usepackage[T1]{fontenc}
\usepackage{amsmath} 
\usepackage{amsthm}
\usepackage{appendix}
\usepackage{mathtools}
\usepackage{etoolbox}
\usepackage{comment}
\usepackage{lipsum}
\usepackage{svg}

\DeclareMathOperator{\E}{\mathbb{E}}

\makeatletter
\def\@email#1#2{%
 \endgroup
 \patchcmd{\titleblock@produce}
  {\frontmatter@RRAPformat}
  {\frontmatter@RRAPformat{\produce@RRAP{*#1\href{mailto:#2}{#2}}}\frontmatter@RRAPformat}
  {}{}
}%
\makeatother
\begin{document}

\preprint{AIP/123-QED}

\title[]{Efficient force field
and energy emulation through  partition of permutationally equivalent atoms}
\author{Hao Li}
 \affiliation{Department of Statistics and Applied Probability, University of California, Santa Barbara, CA 93106, USA}
\author{Musen Zhou}%
\affiliation{ 
Department of Chemical and Environmental Engineering, University of California, Riverside, CA 92521, USA
}%
\author{Jessalyn Sebastian}
    \affiliation{Department of Statistics and Applied Probability, University of California, Santa Barbara, CA 93106, USA}
\author{Jianzhong Wu }
\affiliation{ 
Department of Chemical and Environmental Engineering, University of California, Riverside, CA 92521, USA
}%
\author{Mengyang Gu}
\homepage{Author to whom correspondence should be addressed: \href{mailto:mengyang@pstat.ucsb.edu}{mengyang@pstat.ucsb.edu}
}
\affiliation{Department of Statistics and Applied Probability, University of California, Santa Barbara, CA 93106, USA}

\date{\today}

\begin{abstract}
  Gaussian process (GP) emulator has been used as a surrogate model for predicting force field and molecular potential, to overcome the computational bottleneck of \emph{ab initio} molecular dynamics simulation. Integrating both atomic force and energy in predictions was found to be more accurate than using energy alone, yet it requires  $O((NM)^3)$ computational operations for computing the likelihood function and making predictions, where $N$ is the number of atoms and $M$ is the number of simulated configurations in the training sample, due to the inversion of a large covariance matrix. The high computational cost limits its applications to the simulation of small molecules. The computational challenge of using both gradient information and function values in GPs was recently noticed in machine learning communities, whereas conventional approximation methods may not work well. Here, we introduce a new approach, the atomized force field model, that integrates both force and energy in the emulator with many fewer computational operations. The drastic reduction on computation is achieved by utilizing the naturally sparse covariance structure that satisfies the constraints of the energy conservation and permutation symmetry of atoms. 
The efficient machine learning algorithm extends the limits of its applications on larger molecules under the same computational budget, with nearly no loss of predictive accuracy. 
Furthermore, our approach contains uncertainty assessment of predictions of atomic forces and  energies, useful for developing a sequential design over the chemical input space.

\end{abstract}

\maketitle

\section{INTRODUCTION}

Fast and accurate emulation of atomic forces and energies is essential to access the microscopic details of chemical and biological events via molecular simulation. Classical molecular dynamics (cMD) relies on a pre-defined force field with semi-empirical forms of the potential energy which often lacks accuracy, while \emph{ab initio} MD (AIMD) sacrifices computational efficiency. In principle, machine learning (ML) approaches can provide a surrogate model to achieve both accuracy of AIMD at the computational cost similar to cMD, thus providing new applications that would not be achievable by conventional methods. While recent years have witnessed enormous development of ML potentials, the field is still rapidly evolving. Many theoretical and computational issues remain to be addressed for the efficient representation of potential-energy surfaces.%
Our essential task in establishing an ML force field is to efficiently correlate molecule-level energetics, such as potential energy surface and atomic forces, with the atomic coordinates. 
Deep neural network (DNN) and Gaussian process (GP)
 are popular tools to emulate AIMD simulation containing a large number of single atoms or small molecules (such as H$_2$O)  \cite{bartok2013representing,bartok2018machine,lu202186}. 
Some effective machine learning approaches have been developed to emulate the dynamics of molecules containing a larger number atoms with different types. 
The kernel ridge regression (KRR) approach based on pairwise diatomic positions and nuclear charge, for instance, was proposed to emulate potential energies of organic molecules \cite{rupp2012fast}. Prediction by KRR is equivalent to using the predictive mean in Gaussian process regression (GPR) \cite{wu2020emulating}, where the uncertainty of predictions can be assessed without additional cost in a GPR. 
The Gaussian approximation potential framework (GAP) \cite{bartok2015g}, as another example, approximates the total energy functional through a decomposition of local atomic energy functional by using self-designed atomic neighborhood information. This approach is often used along with the smooth overlap of atomic positions (SOAP) \cite{bartok2013representing}  to measure the local atomic neighborhood information, such that predictions satisfy translational, permutational and rotational symmetries of atoms. The inducing point sparse approximation \cite{snelson2006sparse} is often used to improve computational scalability in these approaches.  
DNN architectures have also been developed to emulate AIMD \cite{gilmer2017neural}, where
a large number of training samples were often used in training the model. The GPR typically requires fewer samples for accurate predictions, because of two reasons. First, GPR is a nonparameteric model and the complexity of the model, such as the number of basis in predictions, increasing with the sample size, which makes it flexible to estimate nonlinear response surface. Second, the predictive mean in GPR has a closed-form expression, and only a few parameters are required to be numerically estimated, whereas DNN typically relies on numerical optimization in a large parameter space. In both approaches, an appropriate descriptor that encodes the geometry information is important for predictions.

 Combining force and energy samples with energy conservation constraints can improve the predictive accuracy of atomic forces and potential energies in AIMD simulation \cite{chmiela2017machine,chmiela2018towards,schutt2018schnet,zhang2018deep,christensen2020fchl}. %
The approach, called gradient-domain machine learning (GDML) \cite{chmiela2017machine}, starts with the conservation of energy. The force on each atom $\mathbf F_i(\mathbf x)$ is related to the potential energy $E(\mathbf x)$ 
\begin{equation}
    {\mathbf F}_i(\mathbf x)=-\nabla_{\mathbf r_i}  E(\mathbf x), 
    \label{equ:grad_E}
\end{equation} where $\mathbf x = [\mathbf r_1,\mathbf r_2,\cdots,\mathbf r_N]$ is a $3\times N$ matrix of atomic Cartesian coordinates for a system with $N$ atoms, and $\mathbf r_i$ denotes the 3D coordinates for each atom. The pairwise inverse distances of atomic positions were often used to construct the descriptor $\mathbf D(\mathbf x)$ of a molecule
\begin{equation}
  \mathbf D(\mathbf x)_{ij} =
    \begin{cases}
      \|\mathbf r_i-\mathbf r_j \|^{-1} & \text{for i  $>$ j},\\
      0 & \text{o.w. },\\
    \end{cases}
    \label{equ:D}
\end{equation} 
where $||.||$ denotes the Euclidean distance. 
Given $M$ configurations of a molecule containing $N$ atoms, parameter estimation and predictions of the atomic forces by the GDML approach \cite{chmiela2017machine} and its symmetric version \cite{chmiela2019sgdml} involve constructing and inverting a $3N M\times 3N M$ Hessian covariance matrix. 
The computational cost of constructing this covariance matrix 
scales as $\mathcal{O} (M^2 N^3)$ 
and the cost of its inversion scales as $\mathcal{O} (M^3 N^3)$, both increasing rapidly along with the number of atoms and the number of  training simulation runs. The large computational cost of the surrogate model prohibits predicting molecular information in larger and more complex systems.

A wide range of approximation methods for alleviating the computational cost of GP models have been proposed in recent years, including, for instance, the induced point approach \cite{snelson2006sparse}, low rank approximation \cite{cressie2008fixed}, covariance tapering  \cite{kaufman2008covariance},  hierarchical nearest neighbor methods \cite{datta2016hierarchical},  stochastic partial differential equation approach \cite{lindgren2011explicit}, and local Gaussian process approach \cite{gramacy2015local}. Although these methods are useful for approximating GP models with observations at a low dimensional input space, none of them focuses on approximating GP models with high-dimensional gradient observations. 
The large computational cost prevents the direct applications of GP models with high-dimensional gradient information in large-scale systems, a problem which was recently realized in the statistics and machine learning communities  \cite{de2021high}. Low rank approximation and sparse approximation of the covariance were studied \cite{eriksson2018scaling}, yet the predictive accuracy can be degraded. The recent approach  \cite{de2021high} reduces the computational complexity for GP with gradient observations with respect to the dimension of gradients, but the method requires $\mathcal O(M^6)$ computational operations, which is prohibitive for moderately large training runs $M$. Surprisingly, we found that after 
enforcing energy conservation and permutation symmetry of atoms onto the covariance function, the covariance matrix of  atomic forces is approximately sparse. 
This property can be utilized to reduce the computational operations substantially
 without sacrificing the accuracy of predictions. 

In this work, we propose a new surrogate model, which is called the atomized force field (AFF) emulator, for predicting the atomic force and the total energy in AIMD simulations. We demonstrate that AFF is computationally more scalable than prior approaches. 
Unlike other sparse GP approximations, the AFF method reduces the complexity of GP with derivatives without sacrificing predictive accuracy, as many terms in the covariance are already close to zero. 
New features of the proposed approach include: First, we partition the atoms into permutationally equivalent (PE) atom sets (formally defined later), where the correlation of atomic forces at different atom sets is found to be almost negligible. Thus, we can decompose the large covariance matrix of the simulated force vectors into small sub-covariance matrices for each permutationally distinguishable atom, where the sub-covariance contains the most information for making predictions. The computational complexity of the AFF emulator is between $\mathcal{O}(NM^3)$ and $\mathcal{O}(N^3 M^3)$, depending on the permutational symmetry of the molecule. For molecules with unique atomic environment on all atoms like uracil, this feature reduces computational operations in matrix inversion from $\mathcal{O}(N^3M^3)$ in the GDML approach \cite{chmiela2017machine} to  $\mathcal{O}(NM^3)$ in the AFF model. For molecules with all identical atoms, the computational complexity is the same as previous approaches. Fortunately, the reduction of complexity is typically more significant on more complex systems, since the size of the largest set of identical atoms is much less than the total number of atoms. For periodic systems such as crystal and metal, the reduction of the computational cost depends on the local atomic structure. 
When the number of PE atoms is close to the number atoms in the unit cell,  the computational reduction is larger. 
Second, inspired by the popular induced input approach in approximating Gaussian processes \cite{snelson2006sparse,wilson2015kernel}, we develop a new model that approximates the prediction of potential energy based on simulated energy and force observations that are strongly correlated to this molecular configuration, which reduces computational operations in emulating potential energy surface.  
Third, 
our model gives both predictions and uncertainty quantification, as any quantile of the predictive distribution has a closed-form expression. Quantifying the uncertainty in predictions is critically important in an inverse problem, such as optimizing molecular structures based on constraints of physical properties. 
Based on these new features, we can accurately predict atomic force vectors and the potential energy from simulation for molecules with more atoms given the same computational budget. 
\begin{figure*}[ht]
    \centering
    \includegraphics[width=17cm,height=6.5cm]{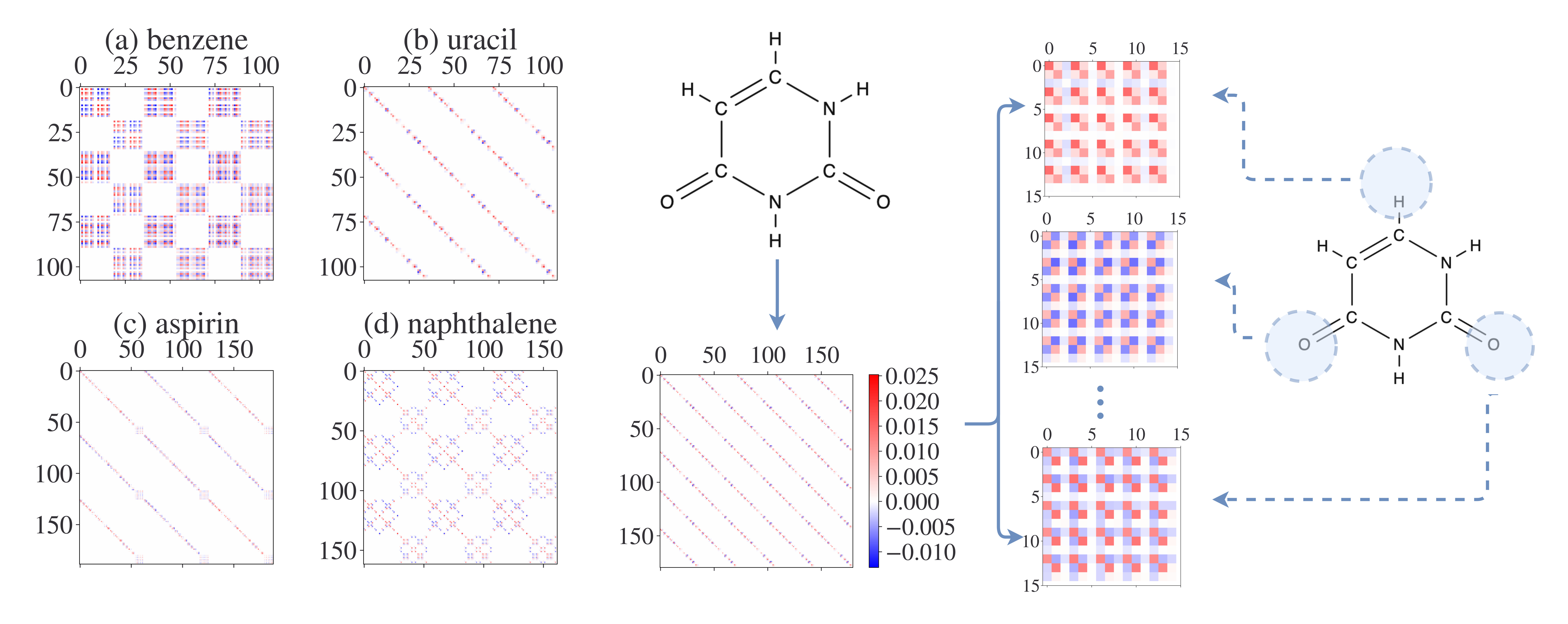}
    \caption{Covariance structure between atomic forces. 
  Displayed on the left are permutationally symmetric covariance matrices of atomic forces on three simulated configurations for (a) benzene, (b) uracil, (c) aspirin and (d) naphthalene. Lighter colors means a smaller absolute covariance, and they indicate most elements in these matrices are near-zero. The right side of the figure shows covariance matrices of atoms in the AFF method on uracil, a molecule for which each atom is its own PE set. Here there is large correlation between atomic force on the same atom at different configurations, but very small correlation between different atoms at the same or different configurations. The rightmost part of the figure shows the subcovariance matrix of atomic force of each atom in uracil across five simulated configurations. }
     \label{fig:AFF_kernel}
\end{figure*}

Furthermore, the AFF model is motivated by a physically informed sparse structure in the covariance matrix and maintains some key physical ingredients of the more computationally intensive approaches, such as the GDML approach and its variants \cite{chmiela2017machine,chmiela2018towards}. In the AFF model, predicting the force on each atom depends on the information of all other atoms in a molecular configuration, expressed as the pairwise distance of atomic positions. Thus, our approach should not be interpreted as a conventional method to capture local atomic information. Empirical results of various examples show that our method is more efficient and accurate than the alternatives based on the same held-out data set.

The rest of the article is organized as follows.  In Sec.  \ref{sec:method},
we first present the motivation of the AFF method.
The notions of PE atoms and an algorithm to find PE atom sets are introduced in Sec. \ref{subsec:TI_set}. 
The atomic force predictions and energy prediction are introduced in  Sec. \ref{subsec:AFF_force} and Sec. \ref{subsec:energy}, respectively. 
In Sec. \ref{sec:result}, we compare our approach with other alternative in  predicting atomic force vectors and potential energies, which demonstrates high predictive accuracy and reliable uncertainty assessment from our approach.  
We conclude this study and provide potential future research directions in Sec. \ref{sec:conclusion}. 

\section{METHODOLOGY}
\label{sec:method} 
Consider a molecule consisting of $N$ atoms.  For two configurations of this molecule, using the vectorized descriptors $\mathbf D(\mathbf x_a)$ and $\mathbf D(\mathbf x_b)$ as input, the covariance of the potential energy of the two configurations is encoded by a kernel function, denoted as $K(\mathbf D(\mathbf x_a), \mathbf D(\mathbf x_b))$. The explicit form of the kernel function will be discussed in Sec. \ref{subsec:AFF_force}.  
By applying the conservation law to energy in Equ. (\ref{equ:grad_E}), the $3N \times 3N$ Hessian covariance matrix of atomic forces, denoted as $\mathbf R(\mathbf x_a, \mathbf x_b)$, has the $(i,j)$th element $(\mathbf R(\mathbf x_a, \mathbf x_b))_{ij} = \nabla_{\mathbf r_{ai}} K(\mathbf D(\mathbf x_a),\mathbf D(\mathbf x_b)) \nabla_{\mathbf r_{bj}}^T$, where $\mathbf r_{ai}$ and  $\mathbf r_{bi}$ denote the $i$th column of $\mathbf x_a$ and the $j$th column of $\mathbf x_b$. Each term of of the Hessian covariance matrix $\mathbf R(\mathbf x_a, \mathbf x_b)$ can be written explicitly by the chain rule discussed in Appendix \ref{app:AFF_correlation}.
Fig. \ref{fig:AFF_kernel} (b) shows the empirical covariance for force vectors of three configurations of uracil in the MD17 dataset \cite{chmiela2017machine}. Note that the correlation between the force vectors of the same atom in three simulations is relatively large, whereas the correlation of force vectors between different atoms is close to zero, which coincides with mathematical results shown in Appendix \ref{app:AFF_correlation}.  
Therefore, we can construct separate force emulators based on the sub-covariance matrix of force vectors of the same atom in different configurations, which reduces  computational complexity.   
Note that, here, we have almost no loss of information, as the covariance contains a sparse structure, which differs from other methods to create a sparse structure to approximate the original covariance matrix \cite{snelson2006sparse}. As seen from numerical comparison in Sec. \ref{sec:result}, our approach is more computationally scalable than the GDML approach and its symmetric variant \cite{chmiela2017machine,chmiela2018towards}.  
Compared with the sparse approximation method \cite{snelson2006sparse},  our approach is an order of magnitude more accurate in terms of out of sample predictive error, shown in the supplementary materials.

Another advantage of our approach comes from incorporating the physical symmetries into the emulator, resulting in higher predictive accuracy. Atoms in a molecule may rotate or switch positions when recorded in simulation, so emulators that encode physical symmetry information can typically achieve higher predictive accuracy  \cite{xie2010permutationally,bartok2013representing,jiang2014permutation,chmiela2018towards,koner2020permutationally}. 
The covariance function of force vectors in the symmetric GDML (sGDML) approach \cite{chmiela2018towards}, for example, was an improved version compared to GDML, as it considers the permutational symmetry of atoms. 
Here, we use a similar way to define  permutational symmetry, and extend it to find the PE groups of atoms. This method identifies a few PE groups for molecules such as benzene, aspirin and naphthalene, in the MD17 dataset \cite{chmiela2017machine}. After grouping these PE atoms and parameterizing the covariance by the permutationally symmetric (PS) kernel function (formally defined in Sec.  \ref{subsec:TI_set}), the model can capture the large absolute covariance between these atoms. This feature empirically improves the predictive accuracy, as will be  shown in Sec. \ref{sec:result}. 

The key idea of the AFF emulator is to partition atoms into different PE groups and to encode large correlation between force vectors of PE atoms at different configurations into the model.   
In the next subsection, we introduce the idea of partitioning the atomic space to obtain PE subsets of atoms.

\subsection{Permutationally equivalent set}
\label{subsec:TI_set}

Because of the existence of  different permutation orders of the atoms for the same molecule, one molecule might have several relevant physical permutation symmetries, leading to the same potential energy surface and force field. A similar idea has been discussed in \cite{chmiela2018towards,chmiela2019towards}. 
To follow this idea, we first define a group of atoms to be permutationally equivalent if they are interchangeable through any permutational operation. It is worthy noting that the permuational symmetry here does not consider the reflection symmetry. 
In other words, atoms are not considered as the same PE set because they may have very different local chemical environments, and thus different forces even if they are plane symmetry. We call atoms from different PE sets permutationally distinct atoms. The AFF approach predicts the force of an atom in a molecule based on the force from its PE group of atoms rather than all atoms in this molecule.  Fig. \ref{fig:AFF_kernel} indicates that we may not need to include all atoms in a large covariance matrix for predicting atomic force to achieve computational efficiency in emulation, as many elements in the Hessian kernel matrix are near-zero. %

For example, all four hydrogen atoms in methane ($CH_{4}$) form one set of PE atoms, while the carbon atom itself is another PE set, as the coordinates of all hydrogen atoms are interchangeable among all permutation symmetries. Benzene ($C_6H_6$), as another example, is comprised of just two sets of PE atoms--the first PE set containing the six carbon atoms and the second PE set containing the six hydrogen atoms. By contrast, all 12 atoms in a uracil molecule ($C_4H_4N_2O_2$) are permutationally distinctive, due to the unique atomic environments of each atom, which leads to 12 PE atom sets.

The PE sets of atoms can be found by minimizing the loss function through the permutation matrix $\mathbf P^*$ \cite{umeyama1988eigendecomposition,chmiela2019sgdml}
\begin{equation}
  \mathbf P^*=\underset{\mathbf P}{\arg\min} \| \mathbf P \mathbf{A_H} \mathbf P^T - \mathbf{A_G} \|,
  \label{equ:permutation}
\end{equation} 
where $\mathbf A_H$ and $\mathbf A_G$ are adjacency matrices of two isomorphic molecules, and $(\mathbf A)_{ij}=\|\mathbf r_i -\mathbf r_j\|$. By analyzing the index location from the permutation matrices of all permutation symmetries on the same type of molecule, we can partition the atoms from the same molecule into sets of PE atoms $ S^i = \{\mathbf r_1^i, \cdots,\mathbf r^i_{l_i} \},$ where $\mathbf r_1^i, \cdots,\mathbf r_{l_i}^i $ are atoms belonging to the $i_{th}$ atom set, for $i=1,...,L$.

Note that the atom in a PE set may exchange positions (e.g. through rotation) and thus the force may be recorded in different order in simulation. The Euclidean distance of the inverse pairwise distance descriptor in Equ. (\ref{equ:D}) cannot capture the similarity between two atomic forces in this scenario.
To represent the large similarity  between force of atoms in a PE set, one may permute the positions of atoms, through a permutationally symmetric (PS) kernel function proposed in \cite{chmiela2019sgdml}: 
\begin{equation}
K_{s}(\mathbf D(\mathbf x_a),\mathbf D(\mathbf x_b))=\frac{1}{S^2}\sum_{p=1}^S \sum_{q=1}^S K(\mathbf D(\mathbf P^*_p \mathbf x_a),\mathbf D(\mathbf P^*_q \mathbf x_b)),
\label{equ:K_sym}
\end{equation}
where $S$ is the number of permutation symmetries found by Equ. (\ref{equ:permutation}), and $\mathbf P^*_p$ is the permutation matrix of $p_{th}$ permutation symmetry. For molecule like uracil, where no permutation symmetry exists, we have $S=1$, $\mathbf P^*_1=\mathbf I_N$, and the kernel reduces to a conventional Hessian kernel. 
The correlation between the $i_{th}$ atom of $\mathbf x_a$ and the $j_{th}$ atom of $\mathbf x_b$ using the PS kernel function are the average of Hessian covariance matrix $(\mathbf R(\mathbf P^*_p \mathbf x_a,\mathbf P^*_q\mathbf x_b))_{ij}$. 

As shown in Fig. 1 (a) and (d), the absolute correlation between the atoms in a PE set  in Equ. (\ref{equ:K_sym}) is much larger than zero.  We found that  using the PE atoms significantly improves the predictive accuracy of atomic forces,  compared with the approach that groups each atom as one set. This result is sensible as 
forces of PE atoms are similar, and the correlation of forces from the PS kernel between PE atoms can capture the similarity. In contrast, the conventional Hessian kernel does not encode the permutational symmetries into the model. 
Note that the correlation of atomic forces from different PE sets is close to zero.This feature allows us to model atomic forces in each PE set separately, which substantially reduces the computational complexity.

\subsection{Atomized force field model}
\label{subsec:AFF_force}

Consider a molecule that has $N$ atoms grouped into $L$ PE atom sets, each set containing $l_i$ atoms, for $i=1,...,L$.  
we decompose the large covariance matrix to construct predictive models for each PE atom set in parallel. Let $\mathbf X=\{\mathbf x_1,...,\mathbf x_M\}$ be $M$ configurations of this molecule that have been simulated from AIMD, and let $\mathbf x^i_j$ be a $3\times l_i$ matrix that contains the $i_{th}$ PE set's atomic coordinates in $\mathbf x_j$. Denote the forces of the atoms of $i_{th}$ PE set in $M$ training configurations by a $3Ml_i$ vector $\mathbf F_i$.  
For a new molecular configuration $\mathbf x^*$, the KRR estimator minimizes the loss function that penalizes both squared error fitting loss, and the complexity of the latent function simultaneous \cite{rasmussen2006gaussian}, leading to a weighted average of the force vectors at $M$ training configurations:   

\begin{equation}
\label{eq:KRR_F}
    \mathbf {\hat F}_i(\mathbf D(\mathbf x^*))= \bm \omega_i^* \mathbf F_i,
\end{equation}
where the weights follow $\bm \omega_i^* = \mathbf R^T_{\mathbf x^*}(\mathbf R+{\lambda}\mathbf I_{3Ml_i} )^{-1}$ with $\mathbf I_{3Ml_i}$ being an identity matrix of size $3Ml_i\times 3Ml_i$. Here $\mathbf R$ is a $3Ml_i \times 3Ml_i$ covariance matrix with the (j,k)th $3l_i \times 3l_i$ block term being the Hessian matrix of the kernel function $\nabla_{\mathbf x_j^{i}} K(\mathbf D(\mathbf x_j),\mathbf D(\mathbf x_k)) \nabla_{\mathbf x_k^{i}}^T$; $\lambda$ is an estimated  regularization parameter;  $\mathbf R_{\mathbf x^{*}}$ is a $3Ml_i \times 3l_i$ matrix, where the $j$th $3l_i \times 3 l_i$ block term is $\nabla_{\mathbf x_j^{i}} K(\mathbf D(\mathbf x_j),\mathbf D(\mathbf x^{*})) \nabla_{\mathbf x^{*i}}^T$.

Note that the KRR estimator in Equ. (\ref{eq:KRR_F}) is the predictive mean of GP regression \cite{rasmussen2006gaussian,kanagawa2018gaussian}. In addition to a point  prediction on an untested run, the GP regression can provide closed-form predictive intervals, which is useful for uncertainty assessment of  predictions. 
Thus, we can construct a GP regression model of atomic forces separately for each PE atom set. Given any $M$ training simulation runs,  the marginal distribution of the force vector $\mathbf F_i$ follows a multivariate normal distribution:
\begin{equation}\label{eq:pp_F}
    \left(\mathbf F_i \mid 
     \mathbf R, \sigma^2_i, \lambda \right) \sim \mathcal{MN}\left( \mathbf 0, \sigma^2_i (\mathbf R+\lambda \mathbf I_{3Ml_i})\right), 
\end{equation} 
for $i=1,\cdots,L$, where 
$\sigma^2_i$ is a variance parameter for the $i$th PE set, and $\lambda$ is the nugget parameter shared across all PE sets. Here the variance parameter $\sigma^2_i$ can differ across different PE sets, as the scale of the force can vary significantly for atoms in each PE atom set. The range and nugget parameter are assumed to be the same, as the smoothness of the latent function that maps atoms' positions to forces are approximately the same across different atom sets. The computational complexity of the predictive mean in a GP emulator with the same kernel and nugget parameters across atom sets is much smaller than the GP emulator with different parameters  \cite{Gu2016PPGaSP}.

The power exponential covariance and the Mat{\'e}rn covariance function are widely used as the covariance function in GP models \cite{rasmussen2006gaussian}. The Mat\'ern kernel function with roughness parameter 5/2 is used as default covariance function of a few GP emulator packages \cite{roustant2012dicekriging,gu2018robustgasp}, as well as the GDML appproach for energy-conserving force field emulation \cite{chmiela2017machine}. This is partly because  the sample path of GP with this kernel is twice differentiable, leading to relatively accurate predictions for both rough and smooth response surfaces.
Here we also use the Mat\'ern kernel function with roughness parameter 5/2:
\begin{equation}
K(\mathbf D(\mathbf x_a),\mathbf D(\mathbf x_b))=\left(1+\sqrt{5} \frac{d}{\gamma}+\frac{5 d^2}{3 \gamma^2}\right) \exp\left(-\sqrt{5}\frac{ d}{\gamma}\right),
\label{equ:K}
\end{equation} 
where $\gamma$ is the range parameter, and $d$ is the Euclidean distance between $\mathbf D(\mathbf x_a)$ and $\mathbf D(\mathbf x_b)$. Similar to the adjustment of the kernel function used in  sGDML \cite{chmiela2018towards},  we transform the Mat{\'e}rn kernel to the PS kernel function in Equ. (\ref{equ:K_sym}) in the AFF emulator, to capture permutational symmetries between PE atoms. 
Conditional on $\gamma$ and $\lambda$, the maximum likelihood estimator (MLE) of $\sigma^2_i$ is $\hat{\sigma}^2_i=S^2_i/ (M 3l_i)$ with $S^2_i=\mathbf F^{T}_i (\mathbf R+\lambda \mathbf I_{3Ml_i})^{-1} \mathbf F_{i}$ for the $i$th PE atom set. The nugget parameter $\lambda$ and the range parameter $\gamma$ can be estimated by numerically optimizing the profile likelihood or by  cross validation with respect to squared error loss in predictions. When the number of training configurations is small, the marginal posterior mode may be used to avoid unstable estimation of the range  and nugget parameters \cite{gu2018jointly}.

Conditional on the estimated parameters $ \hat{\bm \theta}_i=[\hat{\sigma}^2_i,\hat{\gamma},\hat{\lambda}]$, the predictive distribution of the atomic forces in the $i$th PE atom set $\mathbf F_i(\mathbf D(\mathbf x^{*}))$ at any configuration $\mathbf x^{*}$ follows a multivariate normal distribution  \begin{equation}\label{eq:F_multivarite}
\left(
\mathbf F_i(\mathbf D(\mathbf x^{*})) \mid \mathbf F_i, \hat{\bm \theta}_i\right) \sim \mathcal{MN}( \mathbf {\hat F}_i(\mathbf D(\mathbf x^{*})), \hat \sigma^2_i \mathbf K^{*}(\mathbf x^*,\mathbf x^*)),
\end{equation} where the predictive mean vector and predictive covariance matrix follows \begin{align}
    \mathbf {\hat F}_i(\mathbf D(\mathbf x^{*}))
    &=\mathbf R^T_{\mathbf x^{*}}( \mathbf R+\hat{\lambda} \mathbf I_{3Ml_i})^{-1}
    \mathbf F_i, \label{equ:pred_mean} \\
        \hat\sigma^2_i \mathbf K^{*}(\mathbf x^*,\mathbf x^*)&=\hat\sigma^2_i(\mathbf R^{*}  - \mathbf R^T_{\mathbf x^{*}}( \mathbf R+\hat{\lambda} \mathbf I_{3Ml_i})^{-1} \mathbf R_{\mathbf x^{*}}),
\end{align}  
with $\mathbf R^{*}$ being a $3l_i \times 3l_i$  Hessian matrix of the kernel function $\nabla_{\mathbf x^{*i}} K(\mathbf D(\mathbf x^{*}),\mathbf D(\mathbf x^{*})) \nabla_{\mathbf x^{*i}}^T$.

\subsection{Predicting potential energy through the AFF model}
\label{subsec:energy}

Emulating energy based on integrating both simulated force vector and energy can also induce high computational costs, due to computing the inversion of a large covariance matrix of simulated force vectors and energies  \cite{christensen2020fchl}. Here we introduce a computationally feasible approach to emulate the potential energy. 
 For any molecule with atomic configuration  $\mathbf x^*$, 
 the potential energy  $E(\mathbf x^*)$ correlates %
 with the vector of potential energy from previously simulated molecular configurations $\mathbf E=(E(\mathbf x_1),...,E(\mathbf x_M))$, and the unobserved atomic force at this molecular configuration $\mathbf F(\mathbf x^*)$. Conditional on $\mathbf E$ and $\mathbf F(\mathbf x^*)$, the correlation between $E(\mathbf x^*)$  and forces at other configurations is small. 
Thus the predictive distribution of $E(\mathbf x^*)$ conditional on both simulated energy and atomic force $[E(\mathbf x^*) \mid \mathbf E, \mathbf F]$ can be approximated by $(E(\mathbf x^*) \mid \mathbf E, \mathbf F(\mathbf x^*))$, where $\mathbf F(\mathbf x^*)$ can be estimated by the predictive distribution in the AFF model discussed in Section \ref{subsec:AFF_force}. The motivation of the method is relevant to the inducing point approximation approach \cite{snelson2006sparse}, where given outcomes of a function at a set of well-chosen induced pseudo-inputs, the predictive distribution of the outcome at a new input is assumed to be conditionally independent of outputs in the training dataset. Here the inducing input points of $E(\mathbf x^*)$ are $[\mathbf E, \mathbf F(\mathbf x^*)]$, due to large correlation between these variables. Conditional on $(\mathbf E, \mathbf F(\mathbf x^*))$, we assume the  force vector $\mathbf F(\mathbf x^*)$ at this configuration is approximately independent to other training configurations of force vectors. This simplification avoids constructing and computing the large Hessian covariance matrix of force vectors, allowing us to perform inversion of a $(3N+M)\times(3N+M)$ covariance matrix, instead of inversion of a $(3N+1)M\times(3N+1)M$ covariance matrix, in energy prediction. When predicting energy in many new molecular settings,  matrix inversion of the sub-covariance matrix for simulated energy is shared among all predictive distributions. Thus they are only needed to be computed once. Details of efficient computation for predicting energy are discussed in Appendix \ref{app:FastEnergy}. 

For any molecule with atomic coordinates $\mathbf x^*$, the learning objective can be represented by a combined vector of force and energy $\mathbf{EF}=( E(\mathbf x^{*}),\mathbf E^T,\mathbf F^T(\mathbf x^{*}))^T$, where $E(\mathbf x^{*})$ is the potential energy of the molecule with atomic coordinates $\mathbf x^{*}$, 
and $\mathbf F(\mathbf x^{*})$ is the force field vector of this molecule. 
Assuming a GP model for potential energy  with covariance function $K(\cdot,\cdot)$ and mean function  $\mu(\cdot)$, the random vector $\mathbf{EF}$ follows a multivariate normal distribution:
\begin{equation}\label{eq:combine_EF}
    \mathbf{EF}  \sim \mathcal{MN}\left( \bm \mu_{EF}, \bm \Sigma_{EF} \right),
\end{equation}
where the mean vector is given as follows 
\begin{equation*}
    \bm \mu_{EF}=\left( \mu(\mathbf x^{*}), \bm \mu_{\mathbf X}^T, - \nabla_{\mathbf r} \mathbf \mu(\mathbf x^{*})^T   \right)^T, 
\end{equation*}  with $\mu(\mathbf x^*)$ assumed to be an unknown constant $m$ (estimated by MLE in this work), and  $\bm \mu_{\mathbf X}=m\mathbf 1_M$.  The  covariance matrix in Equ. (\ref{eq:combine_EF}) is given as follows
\begin{equation*}
\bm \Sigma_{EF}= \sigma^2 \left(
    \begin{bmatrix}
    K_{\mathbf x^*,\mathbf x^*} & \mathbf K_{\mathbf X,\mathbf x^*} & -\mathbf J^T_{\mathbf x^*,\mathbf x^*}\\
    \mathbf K_{\mathbf x^*,\mathbf X} & \mathbf K_{\mathbf X,\mathbf X} & -\mathbf J_{\mathbf X,\mathbf x^*}\\
    -\mathbf J_{\mathbf x^*,\mathbf x^*} & -\mathbf J_{\mathbf x^*,\mathbf X} & \mathbf R_{\mathbf x^*,\mathbf x^*}
    \end{bmatrix} + \lambda \mathbf I_{M+3N+1}\right),
\end{equation*}
where the upper left four matrix blocks are the covariance of energy vectors $(E(\mathbf x^{*}),\mathbf E^T)^T$. The $(i,j)$ element of the correlation matrix $K_{\mathbf X,\mathbf X}$ is  $K(\mathbf D(\mathbf x_i),\mathbf D(\mathbf x_j))$, for $i=1,...,M$ and $j=1,...,M$, and  $K_{\mathbf x^*,\mathbf x^*}=1$.   The vector $\mathbf K_{\mathbf x^*,\mathbf X}=\mathbf K_{\mathbf x^*,\mathbf X}^T=(K(\mathbf D(\mathbf x^*),\mathbf D(\mathbf x_1)),...,K(\mathbf D(\mathbf x^*),\mathbf D(\mathbf x_M)) )^T$ denotes the correlation between the potential energy at input $\mathbf x^*$ and the potential energy at training inputs $\mathbf X$. In addition, the $3N \times 3N$ correlation matrix between forces is denoted by $\mathbf R(\mathbf x^*,\mathbf x^*)=\nabla_{\mathbf x^{*}} K(\mathbf D(\mathbf x^{*}),\mathbf D(\mathbf x^{*})) \nabla_{\mathbf x^{*}}^T$. Finally, $\mathbf J$ denotes the correlation between energy and forces. Here 
$\mathbf J_{\mathbf x^*,\mathbf X}$ is a $3N\times M$  matrix with the $j$th column being $\nabla_{\mathbf x^*} K(\mathbf D(\mathbf x^*),\mathbf D(\mathbf x_i))$, and $\mathbf J_{\mathbf x^*,\mathbf x^*}$ is the correlation matrix between force and energy for molecule configuration with atom positions $\mathbf x^*$.

Similar to parameter estimation of AFF model discussed in Sec. \ref{subsec:AFF_force}, the mean and variance parameter can be estimated by the MLE of the simulated (training) energy vector below 
\begin{align*}
  \hat{m}&=(\mathbf 1^T_M (\mathbf K_{\mathbf X,\mathbf X}+\lambda \mathbf I_M)^{-1} \mathbf 1_M)^{-1} \mathbf 1^T_M (\mathbf  K_{\mathbf X,\mathbf X}+\lambda \mathbf I_M)^{-1} \mathbf E  \\
  \hat{\sigma}^2&=\frac{\left(\mathbf E-\mathbf 1 \hat{m}\right)^T  (\mathbf K_{\mathbf X,\mathbf X }+\lambda \mathbf I_M)^{-1} \left(\mathbf E-\mathbf 1 \hat{m}\right)}{M}.
\end{align*}
The range parameter $\gamma$ and the nugget parameter $\lambda$  in the kernel function can be estimated through numerical optimization by cross-validation or MLE.

\begin{figure}[ht]
    \centering
    \includegraphics[width=8cm,height=4.7cm]{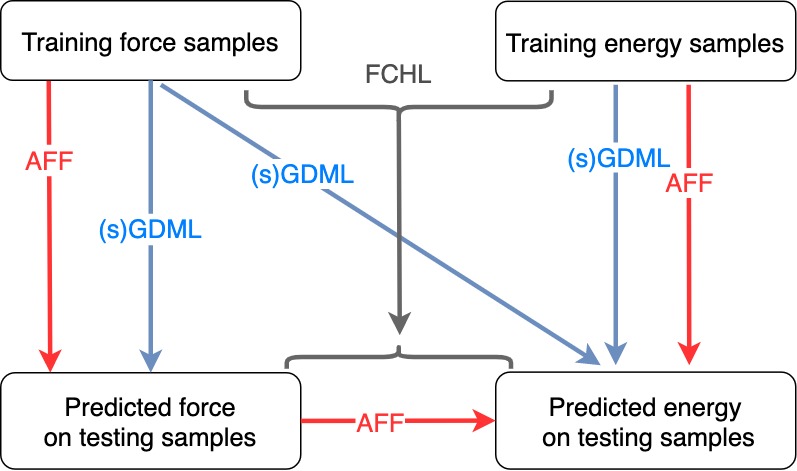}
    \caption{Schematic representation of different approaches in predicting atomic force and potential energy of molecules. GDML and sGDML  methods predict force of molecule at a new configuration based on forces on simulated configurations. The predictive force and energy were used to estimate the energy of this molecule. 
    The FCHL method estimates the energy and atomic force  by a joint model fitted using both the simulated force and energy samples. The AFF method partitions the atoms into PE atoms set and the atomic force of atoms of a new configuration is predicted based on the simulated force of atoms in the same PE set. The energy of molecule at this configuration is predicted based on the predicted atomic force and energy from simulated samples.  
    }
    \label{fig:E_map}
\end{figure}

Based on previous discussion, assuming that the given $[\mathbf E, \mathbf F(\mathbf x^*)]$, $E(\mathbf x^*) $ is approximately independent of the rest of force vectors,  then we have   
\begin{equation}\label{eq:conditional_E}
        \left(E(\mathbf x^{*}) \mid \mathbf E, {\mathbf F},\hat{m},\hat{\sigma}^2, \hat \gamma,\hat \lambda \right) \overset{.}{\sim} \mathcal{MN} \left( \hat E(\mathbf x^*), \hat{\sigma}^2  K_E^*(\mathbf x^*,\mathbf x^*) \right),
    \end{equation}
    where $\overset{.}{\sim}$ denotes the approximation of the predictive distribution, and the predictive mean is a weighted average of training energy $\mathbf E$ and training force $\mathbf F$:
        \begin{equation}
        \hat E(\mathbf x^*)=\bm \omega^*_E \mathbf E+ \bm \omega^*_F  \bf { F}.  
        \end{equation}
    Closed form expressions of 
    $\bm \omega^*_E$, $\bm \omega^*_F$ and $\mathbf K_E^*(\mathbf x^*,\mathbf x^*)$ are derived in Appendix \ref{app:mu_E}.

In practice, the energy on the testing set is estimated in batches using the conditional distribution in Equ. (\ref{eq:conditional_E}). 
The advantage of our method is that we exploit the estimable information from the force vector, but avoid computing the inverse of the gigantic kernel matrix on $\mathbf F$, which substantially simplifies the computation.  

The comparison between the AFF and GDML models for predicting the molecular energy is illustrated in Fig. \ref{fig:E_map}. For GDML, as well as for both sGDML and FCHL 
, all simulated energy and force are used, but the inversion of a large covariance matrix is computationally expensive. Here, conditional on the simulated energy and force of a new molecular configuration, we assume the potential energy of a new molecule is independent of the forces of other molecular configuration simulated before. Since AFF does not need to handle the $3MN \times 3MN$ covariance matrix of  simulated force vectors, it is more scalable for predicting the molecular level information of larger systems. 

\section{NUMERICAL RESULTS}
\label{sec:result}

We evaluate the performance of the AFF approach by analyzing the required training time and learning curves on a variety of molecules, including benzene, uracil, and naphthalene from the MD17 dataset, and aspirin, alpha-glucose, and hexadecane from our simulated dataset (see the supplementary material for molecules in MD17 dataset). We compare the predictive error and required training time from AFF with some of the most commonly used KRR-based models, such as the GDML and sGDML approaches for force and energy predictions. 
All comparisons are implemented under the same training and testing set. In addition, we  provide the uncertainty assessment of  predictions from our model through the proportion of held-out outcomes covered in the $95\%$  predictive interval and the average length of the predictive interval (see Table \ref{table:AFF} for details on the prediction accuracy, required training time, and uncertainty assessment of the AFF predictions). The ratio of the average length of the predictive interval to the range of testing forces $L^{norm}$, and the difference between the 95\% confidence level and the proportion of the held-out samples contained in the predictive interval $\Delta p^{0.95}_{CI} $ on the held-out dataset are shown in Fig.  \ref{fig:UQ}. 
An efficient method should have small predictive error, small training cost, short average length of the predictive interval, and around $95\%$ of the held-out test data covered by the $95\%$ predictive interval. Furthermore, the comparison between our method with the sparse approximation method in a SOAP model  \cite{snelson2006sparse,bartok2013representing} is given in the supplementary material. We do not include them in this section as our method seems to be an order of magnitude more accurate than the method with a sparse approximation to the covariance matrix, in terms of predicting the same held-out dataset.  

\begin{figure*}
  \includegraphics[width=18.5cm,height=7.9cm]{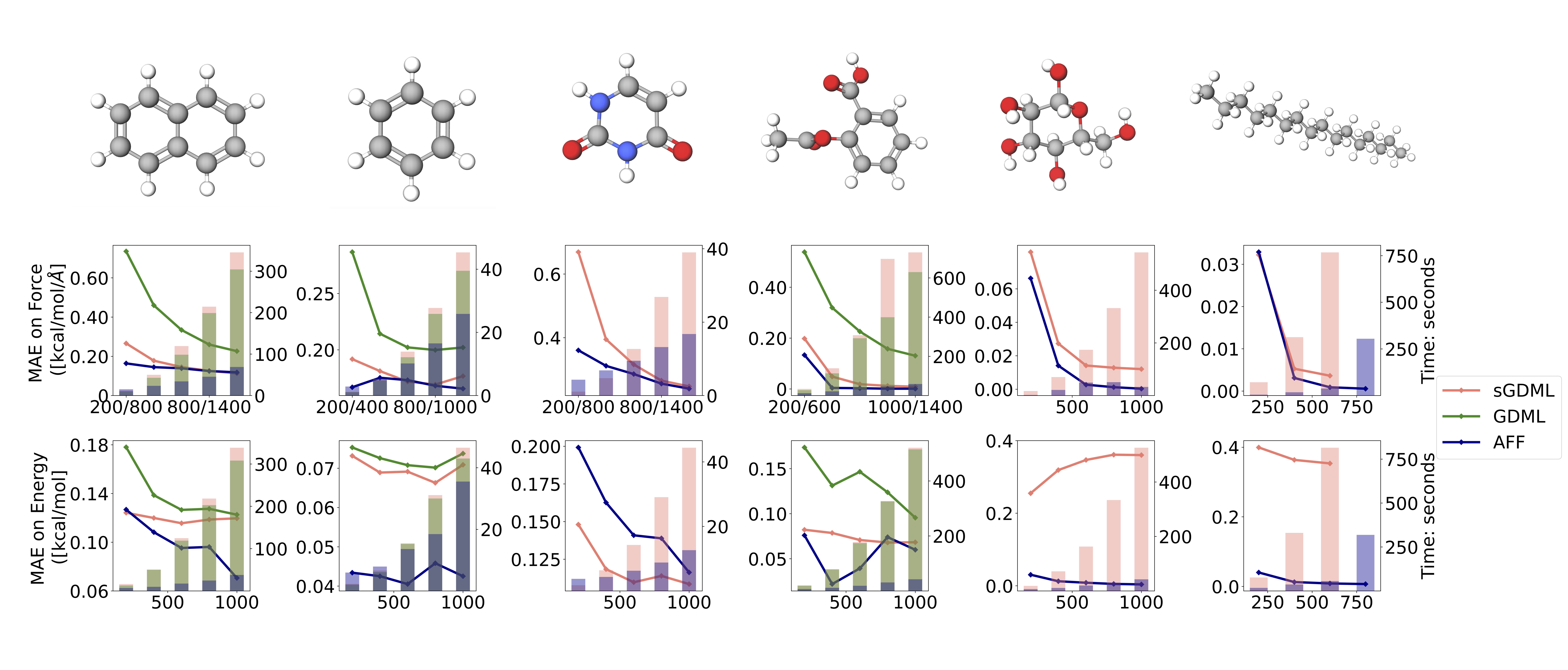}
  \caption{The learning curves for force and energy on naphthalene, benzene, uracil from the MD17 dataset, and aspirin, alpha-glucose, and hexadecane from our simulated dataset (ordered from left to right). Learning curves are presented for the GDML and sGDML methods as well as the AFF method under the same training, validation and testing set. The training sample size (x-axis) is tweaked for AFF method on force prediction. The AFF uses a larger training set for predicting forces for the first four molecules, and the computational time(shown as the blue bars) is still much lower compared with the GDML and sGDML approaches. The top row contains the learning curve [in terms of mean absolute error (MAE)] and training time for the out-of-sample force prediction. The bottom row contains the learning curve and training time for the out-of-sample energy prediction. GDML and sGDML approach are equivalent on uracil, alpha-glucose and hexadecane, as all atoms are permutationally distinct. Thus only two learning curves are shown for those molecules. 
  }
     \label{fig:MAE_all}
\end{figure*}

Previous studies have shown that the GDML and sGDML have relatively small error, compared with other approaches \cite{chmiela2018towards,christensen2020fchl}.  
Indeed, according to Fig. \ref{fig:MAE_all}, the predictive error is relatively small for both approaches.  However, both GDML and sGDML have a large computational cost, mainly due to the inversion of $3NM\times 3NM$ covariance matrix of force vectors at all training configurations.  
Because of the reduced computational order on force prediction by partitioning the atoms into PE atom sets, the AFF model has a smaller predictive error of force prediction (blue curves) when using similar or even less training time (blue histograms) compared to  GDML and sGDML approaches.  
The improved accuracy of force predictions by the AFF model is even more noticeable on the additional simulation of aspirin, alpha-glucose and hexadecane as shown in Fig. \ref{fig:MAE_all}. For some small molecules in the MD17 dataset, the AFF method could achieve better accuracy with less computational cost by using more training samples. For the larger molecules such as glucose and hexadecane, the propose AFF method shows the no loss of accuracy compare to the sGDML predictions at the same number of training samples.

The sGDML approach typically has a smaller predictive error compared with GDML for molecules with at least two PE atom sets, such as  benzene, aspirin and naphthalene molecules, consistent with the result reported in the previous study \cite{chmiela2018towards}. This is because sGDML approach encodes 
the PS kernel  to properly represent the large correlation of forces between atoms in the PE atom set. 
Note that here the reduced computational cost in AFF allows us to train our models with more observations than the GDML and sGDML approaches for predicting the force with a even smaller computational budget. The number of training observations required in training the AFF model, however, is still very small (from a few hundred to a thousand). 

In comparison,  neural network (NN) approaches typically need a larger set of training observations (ranging from $10^4$ to $10^5$) to achieve similar or better predictive performance  \cite{schutt2018schnet,zhang2018deep,zhang2019embedded,unke2019physnet}. Several recent  NN methods are worth exploring \cite{unke2021spookynet,xia2022efficient}, as they seem to require fewer samples than conventional NN approaches. On the other hand,  only two parameters (range and nugget parameters) in the GP model are needed to be numerically optimized, whereas a large number of parameters may need to be numerically optimized in NN approaches.

Given the same number of observations, the error in predicting the potential energy by the AFF model is typically smaller than the sGDML and GDML approaches, as shown in the second rows of Fig.  \ref{fig:MAE_all}. Note that the sGDML with a hybrid loss function \cite{chmiela2019towards} would further improve the accuracy of energy prediction, and we also provide the comparison with it in the supplementary material. 
For some molecules, such as naphthalene and benzene in the MD17 data set, and alpha-glucose and hexadecane in our simulated data set, the AFF model has much smaller predictive error than sGDML approach. This is because our approach incorporates both force and energy vectors in energy prediction when making predictions on energy. Applying the sGDML with the hybrid loss function, the predictive error of it is about the same as the AFF model for alpha-glucose and hexadecane. 
Jointly modeling force and energy was recently studied in \cite{christensen2020fchl}, whereas could have a large computational cost. Here the approximated approach introduced  in Sec.  \ref{subsec:energy} allows us to keep the computational complexity of predicting the energy the same as predicting the force, whereas maintaining relatively high predictive accuracy as that in \cite{christensen2020fchl}. 
For larger molecules, such as alpha-glucose, aspirin, and hexadecane (with 21 - 51 atoms) in our simulated dataset, the computation reduction is huge (see the blue histograms in Fig .\ref{fig:MAE_all}). For these examples, AFF achieves higher accuracy in predicting atomic force, despite costing less than 10\% of the training time of the sGDML approach, as given in Table. \ref{table:AFF}. Furthermore,  the sGDML method requires a larger memory size to storage the covariance of the simulated force vectors. Since we only need to store the covariance of force vectors in each PE atom sets, the memory requirement is often much smaller.

\begin{figure}[t]
    \centering
    \includegraphics[width=9.09cm,height=5.8cm]{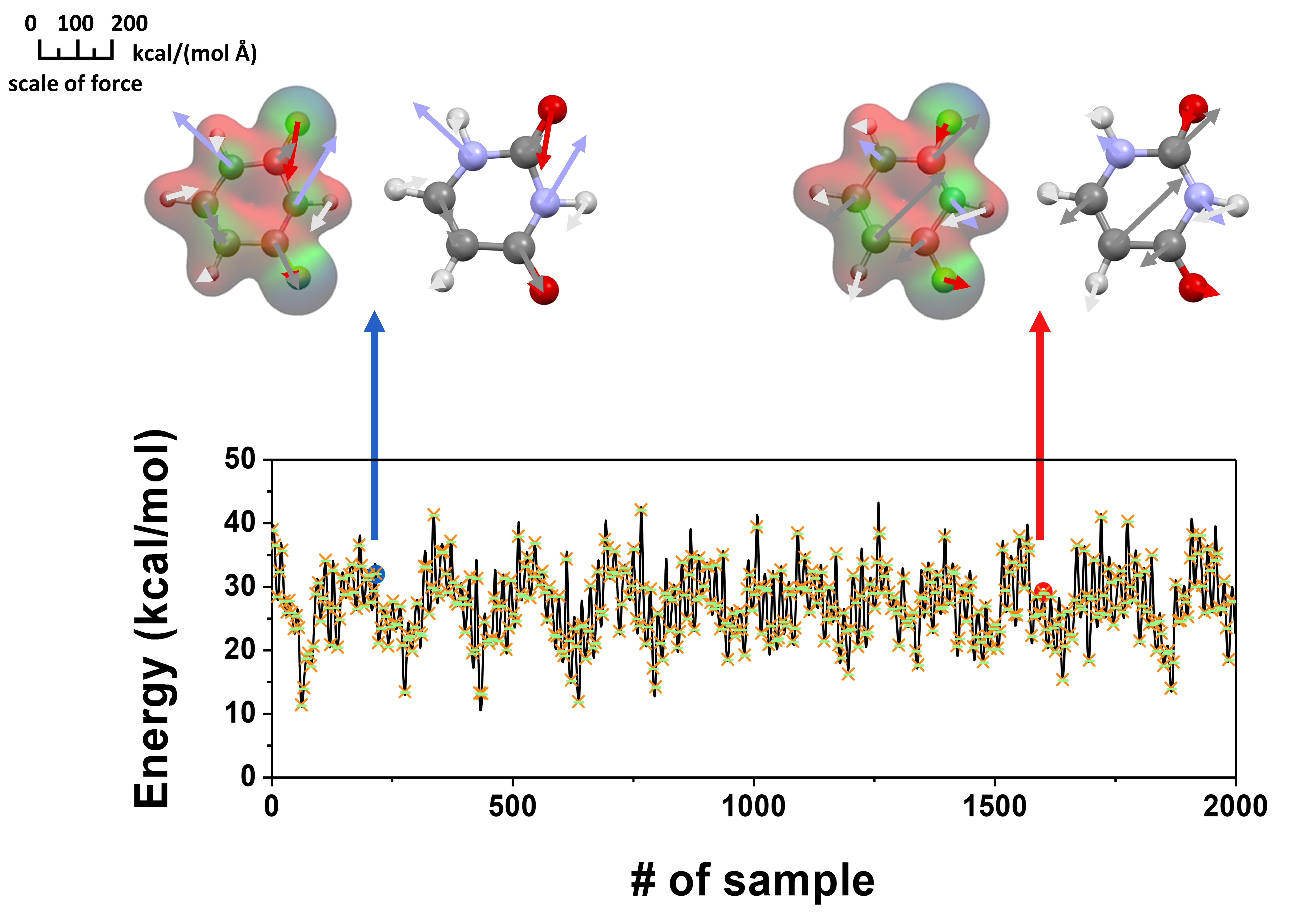}
    \caption{
    Energy of uracil obtained from the AFF method along the AIMD trajectories, where the AIMD energies are displayed in the black curve and predictions of held-out energies from the AFF model are graphed in yellow crosses. The green bars are the 95\% predictive intervals of the AFF energies. As the length of the intervals is very small,  the intervals almost overlap with the AIMD energies and they are nearly invisible. Depicted above the panel of energy in the upper half of the figure is a comparison of AIMD atomic forces and predicted atomic forces by the AFF model on two randomly selected molecules. Within each of the two pairs shown, the same molecule is illustrated twice with the depiction on the left displaying AIMD atomic forces and the depiction on the right displaying the atomic forces by the AFF model. $M=800$ simulated forces and energies were used to train the AFF model for predictions.  
    } 
    \label{fig:Prediction_EF}
\end{figure}

\begin{figure*}[t]
    \centering
    \includegraphics[width=18cm,height=3.75cm]{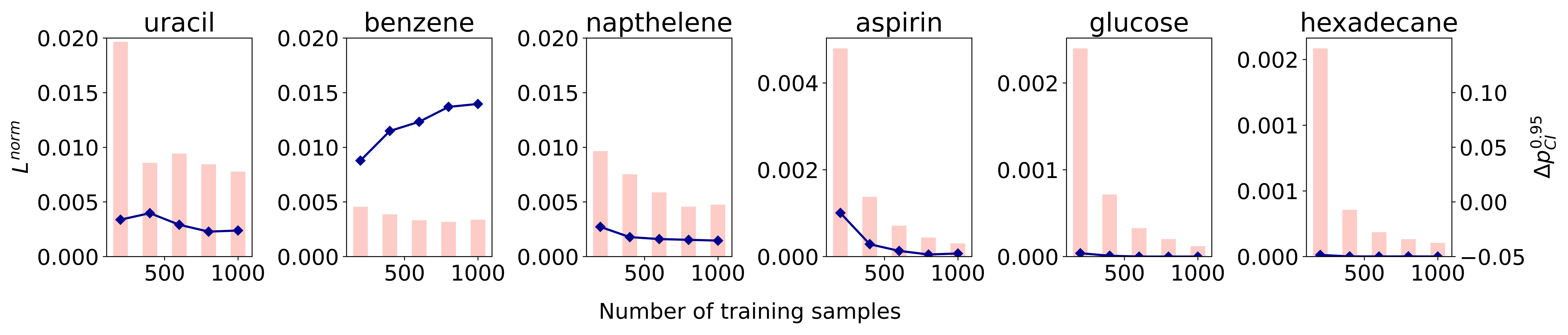}
    \caption{
    The bar charts show the proportion of average AFF predicted atomic force confidence interval $L_{CI}(95\%)$ over the range from testing samples changing with the number of training samples. The blue line charts show the difference $\Delta p $ between confidence level 0.95 and actual coverage of AFF predicted atomic force confidence level.  
    The proportion $L^{norm}$ is given by $\rho =L_{CI}(95\%) / \text{range}(\mathbf F(\mathbf x^*)),$ where 
    $L_{CI}(95\%)=\frac{1}{3N M^*} \sum_{i=1}^{M^*} \sum_{j=1}^{3N} length\{ CI_{ij}(95\%), $ and $ \text{range}(\mathbf F(\mathbf x^*)) = \max(\mathbf F(\mathbf x^*))-\min(\mathbf F(\mathbf x^*))$
    The $\Delta p^{0.95}_{CI} $ is given by $\Delta p^{0.95}_{CI} = 0.95 - P_{CI}(95\%),$ where $P_{CI}(95\%)=\frac{1}{3N M^*} \sum_{i=1}^{M^*} \sum_{j=1}^{3N} 1\{\mathbf F(\mathbf x_i^*)_{j}\} \in CI_{ij}(95\%)$.
    }
   
    \label{fig:UQ}
\end{figure*}

Among all molecules we compared, the AFF model has a larger predictive error for the uracil, using the same number of training input (third panel in the second row in Fig. \ref{fig:MAE_all}).
Since the AFF model estimates the energy based on the emulated atomic force, the accuracy of energy prediction would be reduced when the emulated force is not accurate. As shown in Fig. \ref{fig:MAE_all}, the estimated force by AFF for uracil is not accurate when the number of the training sample is small.  
This problem can be solved by using a moderately large sample size ($\approx$ 1000) to achieve similarly accurate predictions as the sGDML model.  
The predictive energy vector by the AFF model along the AIMD trajectories is graphed  in Figure \ref{fig:Prediction_EF} along with the held-out energy in the simulation. Also plotted are the predictive atomic forces and the truth at two held-out configurations. Based on $M=800$ simulated forces and energies, predictions of potential energies and forces by the AFF model are accurate.

\begin{table*}
\caption{The second and third column show the MAE on estimated energy and force. The fourth column is the training time of the model at shown force accuracy,  which is provided in seconds.  The numbers in parentheses are the sGDML results, and they are tested under same held-out test set. The specific criteria employed are the following: $
    P_{CI}(95\%)=\frac{1}{3N M^*} \sum_{i=1}^{M^*} \sum_{j=1}^{3N} 1_\{\mathbf F(\mathbf x_i^*)_{j}\} \in CI_{ij}(95\%)$, $
    L_{CI}(95\%)=\frac{1}{3N M^*} \sum_{i=1}^{M^*} \sum_{j=1}^{3N} \mbox{length}\{ CI_{ij}(95\%)\},
$ where $M^*$ is the number of test samples, $\mathbf F(\mathbf x_i^*)_{j}$ is the $j_{th}$ element from the force vector prediction of the output of the $i_{th}$ held-out molecule; $CI_{ij}(95\%)$ is the 95\% predictive credible interval from the multivariate normal distribution in  (\ref{eq:F_multivarite}); and $length\{ CI_{ij}(95\%)\}$ is the average length of the 95\% predictive credible interval. The number of training samples used in AFF and sGDML (in parentheses) method is: naphthalene 1600 (1000), benzene 1200 (1000), uracil 1600 (1000), alpha-glucose 1000 (1000), aspirin 1200 (1000), and hexadecane 600 (600). 
}

\begin{ruledtabular}
\begin{tabular}{cccccc}
     \multicolumn{6}{c}{Performance of AFF} 
     
     \\
     \hline
     Molecule    & Energy [kcal/mol] & Force [kcal/mol/\r{A}]& Training Time [s] & $P_{CI}$ (95\%) & $L_{CI}$ (95\%)\\
     \hline 

    Naphthalene   & 0.07 (0.12)  & 0.11 (0.11) & 68 (345) & 98.5\% & 1.36 \\
      \hline
    Benzene    & 0.04 (0.07) & 0.173 (0.176) & 23 (45) & 85\%  & 0.7 \\
      \hline
    Uracil    & 0.10 (0.10) & 0.239 (0.249) &  16 (43) & 97.8\%  & 2.37\\ 
      \hline
    Alpha-glucose    & 0.09 (0.36) & 0.0003 (0.012) & 32 (543) & 100\% & 0.03\\   
      \hline
    Hexadecane   & 0.008 (0.35)& 0.0008 (0.003) &  37 (767) & 99\% & 0.05 \\   
      \hline
    Aspirin   & 0.06 (0.09) & 0.0028 (0.009) & 32 (629) & 99.8\% & 0.08\\
\end{tabular}
\end{ruledtabular}
\end{table*}

Table \ref{table:AFF} gives the predictive error of force vectors and energy, the percentage of forces covered in the predictive interval, the average length of the predictive intervals of forces and computational costs in emulation from different methods. First, the predictive error of AFF methods for both forces and energy is typically not larger than the sGDML approach.  For some molecules such as alpha-glucose and hexadecane, the predictive error of the AFF model seems to be one order of magnitude smaller, based on the same held-out test set. It is worth noting that it takes the AFF model less  computational costs (ranging from 1/2 to 1/30 of costs compared to sGDML) to achieve a similar or higher level of predictive accuracy. These results indicate the AFF model is more efficient in emulating atomic forces and energy in AIMD simulation. Furthermore, around $95\%$ (or higher percentage) of the held-out atomic forces are covered by relatively short  $95\%$ predictive intervals from the AFF approach, indicating that the AFF model provides a reliable way to quantify the uncertainty in predictions.

Furthermore, it is worth mentioning that the reduction of computational cost by the AFF model is more pronounced on molecules with more PE atoms' sets, such as alpha-glucose, aspirin, and hexadecane. For molecules with fewer PE sets such as benzene (where we can only partition the atoms into two PE sets for each configurations), the computational reduction will be smaller. Thus, our approach may be useful for reducing the computational cost of interactions between a large number of molecules, as most PE sets may only contain one atom. 

Finally, uncertainty assessments of predictions of the AFF approach are shown in  Fig. \ref{fig:UQ}. Compared with the range of observations, the average length of $95\%$ (pink bar) is much shorter,  indicating small uncertainty associated with predictions. The difference between the number of held-out test samples covered $95\%$ interval and the nominal $95\%$ range (blue curves) is small, meaning that the uncertainty is accurately quantified. The internal assessment of the uncertainty of the AFF model can be used to identify the input region with large uncertainty, and sequentially design simulation runs for uncertainty reduction or Bayesian optimization \cite{snoek2012practical,shahriari2015taking}.

\section{Concluding remarks}
\label{sec:conclusion}
We have proposed an accurate and computationally efficient approach to predict potential energy surfaces and molecular force fields in \emph{ab initio} simulation. While the theoretical framework of the gradient-based KRR and GP models such as GDML, sGDML, and FCHL, was already established, the challenge posed by the huge computational cost limited the applicability of these methods in emulating systems with a larger size of molecules. We propose the AFF emulator to overcome this computational challenge without compromising its accuracy. The efficient emulation of forces was hinged upon the fact that the similarity of atomic forces between permutationally equivalent atoms is high, whereas the correlation is small across different permutationally equivalent atom sets. By partitioning the atoms of a molecule into different atom sets, the AFF model can capture a large correlation of forces between PE atoms, thereby providing accurate predictions of atomic forces of the molecule at a new configuration with less computational cost. Second, we developed a new  approach to reduce the computational complexity for emulating the potential energy, compared to a joint model of energy and atomic forces of simulated configurations. Numerical results have shown predictions by the AFF emulator are more accurate than alternative approaches, given the same computational budget. The AFF approach discovers a novel path on representing a correlation between forces and energy with significantly lower computational cost.

There are a few potential research directions for emulating simulations of a large-scale system with interactions between a large number of atoms. First, for emulating simulation involving a large number of molecules, one may represent interactions between atoms by partitioning the atoms into PE atoms to decompose the covariance matrix for efficient computation. Second, sparse Cholesky factorization or Markov models may be used to reduce the large computational cost when the required number of simulations used in training machine learning models gets large. Third, given a set of physical constraints, one may inversely design the atomic positions to achieve a particular force field, or potential energy. The uncertainty of the map from atomic positions to potential energy learned by the AFF model is important for a sequential design to reduce the uncertainty of this problem.  
\section*{Supplementary material}
The supplementary material contains numerical comparison between AFF models and sparse approximation approaches, and prediction results on additional molecules.  
\section*{Acknowledgements}
 This study was supported by the U.S. National Science Foundation's Computational and Data-Enabled Science and Engineering Program under Award No. 2053423. 
MZ and JW acknowledge financial support from the U.S. National Science Foundation’s Harnessing the Data Revolution (HDR) Big Ideas Program under Grant no. NSF 1940118. 

\section*{Data Availability}
The data and the code that support the findings
of this study openly available in GitHub at \url{https://github.com/UncertaintyQuantification/AFF} 

\appendix
\section{The correlation of force between different atoms}
\label{app:AFF_correlation}
The correlation between the force of two molecules $a$ and $b$ with positions $\mathbf x_a$ and $\mathbf x_b$ is a $3N \times 3N$ matrix $\mathbf R(\mathbf x_a, \mathbf x_b)$, where the $(i,j)$ element follows $(\mathbf R(\mathbf x_a, \mathbf x_b))_{ij} = \nabla_{\mathbf r^{ai}} K(\mathbf D(\mathbf x_a),\mathbf D(\mathbf x_b)) \nabla_{\mathbf r^{bj}}^T$. Direct computation through the chain rule gives $\nabla_{\mathbf r^{ai}} K(\mathbf D(\mathbf x_a),\mathbf D(\mathbf x_b)) = \sum_{p,q=1}^{N} \frac{\partial \mathbf D(\mathbf x_a)_{pq}}{\partial \mathbf r^{ai}} \frac{\partial K(\mathbf D(\mathbf x_a),\mathbf D(\mathbf x_b)}{\partial \mathbf D(\mathbf x_a)_{pq}} $. 
Based on the form of the descriptor matrix $\mathbf D(\mathbf x)$ in Equ. (\ref{equ:D}), 
the gradient of $(p,q)$ element of $\mathbf D(\mathbf x)$ w.r.t $\mathbf r_i$ follows
\begin{equation}\label{eq:grad_D}
  \frac{\partial \mathbf D(\mathbf x)_{pq}}{\partial \mathbf r_i} =
    \begin{cases}
      -\frac{\mathbf r_p -\mathbf r_q}{\| \mathbf r_p -\mathbf r_q\|^3} & \text{$p>q$ and $i=p$},\\
      \frac{\mathbf r_p -\mathbf r_q}{\| \mathbf r_p -\mathbf r_q\|^3} & \text{$p>q$ and $i=q$},\\
      0 & \text{o.w }.\\
    \end{cases}       
\end{equation} The correlation between the $i$th atom of molecule $a$ and the $j$th atom of molecule $b$ is given by
\begin{align*}
    &(\mathbf R(\mathbf x_a, \mathbf x_b))_{ij} = \nabla_{\mathbf r_{ai}} K(\mathbf D(\mathbf x_a),\mathbf D(\mathbf x_b)) \nabla_{\mathbf r_{bj}}^T \\
    &= \sum_{pq=11}^{NN} \sum_{mn=11}^{NN} \frac{\partial^2 K}{\partial \mathbf D_{pq} \partial \mathbf D_{mn}} \frac{\partial \mathbf D(\mathbf x_a)_{pq}}{\partial \mathbf r_{ai}} \frac{\partial \mathbf D(\mathbf x_b)_{mn}}{\partial \mathbf r_{bj}}\\
    &= \begin{cases}
    \frac{\partial^2 K}{\partial \mathbf D_{ij} \partial \mathbf D_{ij}} \frac{\partial \mathbf D(\mathbf x_a)_{ij}}{\partial \mathbf r_{ai}}  \frac{\partial \mathbf D(\mathbf x_b)_{ij}}{\partial \mathbf r_{bj}} & \text{if $i>j$},\\
    \frac{\partial^2 K}{\partial \mathbf D_{ji} \partial \mathbf D_{ji}} \frac{\partial \mathbf D(\mathbf x_a)_{ji}}{\partial \mathbf r_{ai}} \frac{\partial \mathbf D(\mathbf x_b)_{ji}}{\partial \mathbf r_{bj}}   & \text{if $i<j$},\\
    \underset{\text{p or q=i; } }{\sum}\underset{\text{m or n=i}}{\sum} \frac{\partial^2 K}{\partial \mathbf D_{pq} \partial \mathbf D_{mn}}\frac{\partial \mathbf D(\mathbf x_a)_{pq}}{\partial \mathbf r_{ai}} \frac{\partial \mathbf D(\mathbf x_b)_{mn}}{\partial \mathbf r_{bj}} & \text{if $i=j$},\\
    \end{cases}
\end{align*} 
where $\frac{\partial^2 K}{\partial \mathbf D_{pq} \partial \mathbf D_{mn}}$ is the simplified notation of $\frac{\partial^2 K(\mathbf D(\mathbf x_a),\mathbf D(\mathbf x_b))}{\partial  D(\mathbf x_a)_{pq} \partial \mathbf D(\mathbf x_b)_{mn}}.$

According to the above equation,  when $i=j$, the number of terms in the summation is much larger than the number of terms when $i\neq j$, which indicates that the absolute correlation  between different atoms  is typically smaller than the correlation between same atoms. This empirical result typically holds for usual kernel functions such as Gaussian kernel and Mat{\'e}rn kernel before enforcing the constraint for permutational symmetry.

After applying the PS kernel function from Equ. (\ref{equ:K_sym}), the absolute correlation between different PE atoms is typically smaller than the correlation between same PE atoms. 
As shown in e.g. part (a) in Fig. \ref{fig:AFF_kernel} for benzene, after adopting the KS kernel function, 
the correlation  between PE atoms is large due to permutational symmetries, whereas the correlation of force between atoms in different PE atom sets is still small, indicating that we may only need to condition on forces of atoms in the same PE atom set for efficiently calculating the predictive distribution. 

\section{Fast Predictions of potential energies in batches}
\label{app:FastEnergy}
This section discusses the efficient way to calculate the inversion of the covariance matrix in the AFF model on energy prediction. We achieve the reduced computational cost by predicting the molecules' energies in batches rather than the energy for one configuration each time. 
Let $b$ be the batch size, and $\mathbf x^{b*}=[\mathbf x_1^*,\cdots,\mathbf x_b^*]$ be $b$ configurations of a molecular structure. For predicting the energy of a new configuration of this molecular structure, we need to calculate the inversion of $\mathbf \Sigma_{sub}$, where $\mathbf \Sigma_{sub}=\begin{bmatrix}
   \mathbf A & \mathbf B\\
   \mathbf C & \mathbf D
\end{bmatrix}$, with $\mathbf A=\mathbf K_{\mathbf X,\mathbf X}+\lambda \mathbf I_{M}$ being a $M \times M$ matrix,  $\mathbf B=-\mathbf J_{\mathbf X, \mathbf x^{b*}} $ being a $M \times 3Nb$ matrix, $\mathbf C=-\mathbf J_{\mathbf x^{b*}, \mathbf X}$ being a $3Nb \times M$ matrix, and $\mathbf D= \mathbf R_{\mathbf x^{b*},\mathbf x^{b*}}+\lambda \mathbf I_{3Nb}$ being a $3Nb \times 3Nb$ matrix. 
Note that the sub-covariance matrix of training energy samples $\mathbf A$ is the same among all different batches in prediction. Thus we need to invert  $\mathbf A$ once and by applying the block matrix inversion for $\bm \Sigma_{sub}$, we have: \begin{align}
    \mathbf \Sigma^{-1}_{sub} = \begin{bmatrix}
       \mathbf{A}^{-1}+\mathbf{A}^{-1} \mathbf{B} \mathbf D^{*-1} \mathbf{C A}^{-1} & -\mathbf{A}^{-1} \mathbf{B}\mathbf D^{*-1} \\
-\mathbf D^{*-1} \mathbf{C A}^{-1} & \mathbf D^{*-1}
    \end{bmatrix},
\end{align} where $\mathbf D^*=\mathbf{D}-\mathbf{C A}^{-1} \mathbf{B}$. Accordingly, for each batch samples, we just need to do a matrix inversion on a $3Nb \times 3Nb$ matrix $\mathbf D^*$, which is much faster than inverting the entire covariance matrix $\bm \Sigma_{sub}$ for each batch sample.

\section{Predictive distribution of potential energy}
\label{app:mu_E}
To simplify the notation, we would use the batch size $b=1$ in this section. 
Conditional on simulated energies in the training dataset $\mathbf E$, and the atomic force $\mathbf F(\mathbf x^*)$ at the new configuration $\mathbf x^*$, and the estimated parameters $\hat{\bm \theta}=[\hat m,\hat{\sigma}^2,\hat{\gamma},\hat{\lambda}]$, the conditional distribution of the energy at this configuration follows 
\begin{equation}E(\mathbf x^{*}) \mid \mathbf E, \mathbf F(\mathbf x^*), \hat{\bm \theta} \sim \mathcal{MN}\left(\mu^*_E(\mathbf x^*), \hat{\sigma}^2  K_E^{**}(\mathbf x^*,\mathbf x^*)\right),
\label{equ:cond_dist_F}
\end{equation}
where the conditional mean follows
\begin{equation*}
    \mu^*_E(\mathbf x^*)=\hat{m}+\begin{bmatrix}
    \mathbf K_{\mathbf x^*,\mathbf X} & -\mathbf J^T_{\mathbf x^*,\mathbf x^*}
    \end{bmatrix} \bm  \Sigma_{sub}^{-1}
    \begin{bmatrix}
        \mathbf E-\hat{m} \mathbf 1_M \\
        \mathbf F(\mathbf x^*)
    \end{bmatrix},
\end{equation*} with $ \mathbf \Sigma_{sub}=\begin{bmatrix}
    \mathbf K_{\mathbf X,\mathbf X}& -\mathbf J_{\mathbf X, \mathbf x^*} \\
    -\mathbf J_{\mathbf x^*, \mathbf X} & \mathbf R_{\mathbf x^*,\mathbf x^*}
\end{bmatrix}+\lambda \mathbf I_{3N+M},$ and the conditional variance follows  \begin{equation*}
\mathbf K_E^{**}(\mathbf x^*,\mathbf x^*)= K_{\mathbf x^*,\mathbf x^*}- \begin{bmatrix}
    \mathbf K_{\mathbf x^*,\mathbf X} & -\mathbf J^T_{\mathbf x^*,\mathbf x^*}
    \end{bmatrix}  \bm  \Sigma_{sub}^{-1} \begin{bmatrix}
    \mathbf K^T_{\mathbf x^*,\mathbf X} \\ -\mathbf J_{\mathbf x^*,\mathbf x^*}
    \end{bmatrix}.\end{equation*} 

Note that we do not observe $\mathbf F(\mathbf x^*)$  and thus Equ. (\ref{equ:cond_dist_F}) cannot be directly used for predicting energy at the new configuration $\mathbf x^*$. 
Given the energy of training set $\mathbf E$ and the atomic force of training set $\mathbf F$, we use the total expectation to integrate the unobserved force vector by its predictive distribution: 
\begin{align*}
    \hat E(\mathbf x^*) &=\E[E(\mathbf x^{*}) \mid \mathbf E, \mathbf F] \\
    &=\E[\E[ E(\mathbf x^{*}) \mid \mathbf E, \mathbf F, \mathbf F(\mathbf x^*)]]  \\ 
    &\overset{.}{=} \E[\E[ E(\mathbf x^{*}) \mid \mathbf E,  \mathbf F(\mathbf x^*)]\mid \mathbf F]\\
    &=\E[\mu^*_E(\mathbf x^*) \mid \mathbf E, \mathbf F],
    \end{align*}
    where $\overset{.}{=}$ denotes the approximation of 
    $\E[ E(\mathbf x^{*}) \mid \mathbf E, \mathbf F, \mathbf F(\mathbf x^*)]$ by $\E[ E(\mathbf x^{*}) | \mathbf E, \mathbf F(\mathbf x^*)] $, which is equivalent to assume that given $(\mathbf E, \mathbf F(\mathbf x^*))$, $E(\mathbf x^*) $ is independent of the rest of force vectors in simulated configurations. 
Plugging the predictive mean $\hat{\mathbf F}(\mathbf x^*)$ from the above equation to replace $\mathbf F(\mathbf x^*)$ in $\mu^*_E(\mathbf x^*)$, we  approximate the predictive mean of energy for $ \E[{\hat{E}(\mathbf x^*)} \mid \mathbf E, \mathbf F]$ by  $\hat E(\mathbf x^*)$ with the following expression:  
\begin{equation*}
    \hat E(\mathbf x^*) {=} \hat{m}+ \begin{bmatrix}
    \mathbf K_{\mathbf x^*,\mathbf X} & -\mathbf J^T_{\mathbf x^*,\mathbf x^*}
    \end{bmatrix}
      \bm \Sigma_{sub}^{-1}
    \begin{bmatrix}
        \mathbf E-\hat{m} \mathbf 1_M \\
        \hat{\mathbf F}(\mathbf x^*) \\
    \end{bmatrix}.
\end{equation*} 
The predictive variance in Equ. (\ref{eq:conditional_E}) can be computed by properties of multivariate normal distributions: 
 \begin{equation*}
     K^*_E(\mathbf x^*,\mathbf x^*)= K_{\mathbf x^*,\mathbf x^*}- \begin{bmatrix}
    \mathbf K_{\mathbf x^*,\mathbf X} & -\mathbf J^T_{\mathbf x^*,\mathbf x^*}
    \end{bmatrix}  \bm  \Sigma_{sub}^{-1} \begin{bmatrix}
    \mathbf K^T_{\mathbf x^*,\mathbf X} \\ -\mathbf J_{\mathbf x^*,\mathbf x^*}
    \end{bmatrix}.
\end{equation*}

Let $\mathbf W_{1}$ and $\mathbf W_{2}$ be the first $1 \times M$ block matrix and the latter $1 \times 3N$ block matrix  of $ \begin{bmatrix}
    \mathbf K_{\mathbf x^*,\mathbf X} & -\mathbf J^T_{\mathbf x^*,\mathbf x^*}
    \end{bmatrix}
    \bm  \Sigma_{sub}^{-1},$ 
    respectively,
    and $\hat{\mathbf F}(\mathbf x^*)=\bm \omega_F \mathbf F$, where $\bm \omega_F$ follows from
    Equ. (\ref{eq:KRR_F}). 
    We can also write the $\hat E(\mathbf x^*)$ as the weighted average value of $\mathbf E$ and $ \hat{\mathbf F}(\mathbf x^*)$: \begin{equation}
\hat E(\mathbf x^*)=\bm \omega^*_E \mathbf E+ \bm \omega^*_F {\mathbf F},
\end{equation} where \[\bm \omega^*_E = (1- \mathbf W_1 \mathbf 1_M) (\mathbf 1^T_M \mathbf K_{\mathbf X,\mathbf X}^{-1} \mathbf 1_M)^{-1} \mathbf 1^T_M \mathbf \mathbf K_{\mathbf X,\mathbf X}^{-1} + \mathbf W_1,\] and $
 \bm \omega^*_F = \mathbf W_2 \bm \omega_F$.\\

\section{Simulation details}
\label{app:simulation_detail}
In addition to molecules available from the MD17 dataset, force and energy of additional molecules with more atoms and complicated structure are generated in this work. \emph{ab initio} MD (AIMD) simulation is performed via Q-Chem to generate highly accurate molecular force and energy to benchmark AFF and other machine learning force field. In this work, all AIMD simulations are carried in NVT ensemble with timestep of 1 fs at room temperature (300 K). All the calculations were performed at the level of Perdew-Burke-Ernzerhof(PBE)/6-31G(d,p). vdW interactions are taken into account by using TS-vdW method. The Nosé-Hoover thermostat is used to control the temperature.

\section{Timings}
All timings were performed on a compute cluster equipped with Intel 6148 CPUs (20 cores each) with a high speed OmniPath interconnect. The compute nodes consist of 64 nodes of 40 core/192GB of RAM compute systems, 4 nodes with 768GB of RAM plus 300 GB Intel Optane Memory Drive, and 3 GPU nodes with four NVIDIA V100/32GB GPUs with NVLINK.

\thispagestyle{empty}

\renewcommand\refname{References}
\renewcommand{\thepage}{}

\bibliography{References_2020}

\end{document}